\documentclass[showpacs,superscriptaddress,floatfix,twocolumn,prl]{revtex4}

\usepackage{amsmath}
\usepackage[dvips]{graphicx}
\usepackage{epsfig}
\usepackage{tabularx}

\begin{document}

\title{Chaotically spiking attractors in suspended mirror optical cavities}
\author{Francesco~Marino and Francesco~Marin}
\affiliation{Dipartimento di Fisica, Universit\`a di Firenze, INFN
Sezione di Firenze, and LENS,\\ Via Sansone 1, I-50019 Sesto Fiorentino
(FI), Italy}

\date{\today}
\begin{abstract}
A high-finesse suspended mirror Fabry-Perot cavity is experimentally studied in a regime where radiation pressure and photothermal effect are both relevant. The competition between these phenomena, operating at different time scales, produces unobserved dynamical scenarios where an initial Hopf instability is followed by the birth of small-amplitude chaotic attractors which erratically but deterministically trigger optical spikes. The observed dynamical regimes are well reproduced by a detailed physical model of the system.
\end{abstract}


\maketitle

In the last years, suspended mirror resonators have become a research hot topic. In such systems, the intracavity field and the mirror motion are coupled through radiation pressure. In the strong coupling regime they could enable the generation of non-classical states of light, quantum nondemolition measurements of the field quadratures, the creation of entangled states of light and mirrors and eventually the observation of the quantum ground state of a macroscopic mechanical oscillator (for a review of these topics, see Refs. \cite{qgs1,qgs2}). However, even for lower light powers, the action of radiation pressure and photothermal effect (i.e., thermal expansion of the mirrors due to the absorbed intracavity light) become highly nonlinear.
In particular, radiation pressure nonlinearity produces bistability (as shown in the seminal experimental works of Refs. \cite{walther,gozzini}) and, thanks also to the delay of the electric field build-up in the cavity, regenerative oscillations that eventually become chaotic \cite{in6,in2,in7}. Similar phenomena are generated by thermal effects such as photothermal expansion or bolometric force as shown, for instance, in Ref. \cite{in8}. Therefore the interaction between the two effects, that can both be relevant in quantum opto-mechanical experiments, is expected to drive the system in regimes of extraordinary complexity, that can rise even without the electric field retardation \cite{marino2007}.
Indeed, recent theoretical works \cite{noi1,noi} predict the birth of novel kinds of dynamics  including spontaneous generation of optical spikes.

Irregular spikes sequences are crucial phenomena in the kinetics of chemical reactions \cite{zha,koper} and neural dynamics \cite{ex1,mainen}. While in two-dimensional systems they are necessarily related to the presence of noise, in higher dimensions they can arise from complicated bifurcation cascades, thus having a \emph{purely deterministic} origin. Typical time traces consist of large pulses separated by irregular time intervals where the system displays small-amplitude chaotic oscillations (chaotic spiking).

In many cases ~\cite{chemical,plasma,arecchi}, these phenomena can be interpreted in terms of the well known Shilnikov homoclinic chaos ~\cite{shil}. However, Ref. \cite{noi} proposes that chaotic spiking attractors could be observed also in excitable systems governed by one slow and at least two fast variables. Excitable systems possess a stable attractor (rest state) which can be forced to spike by stimuli of amplitude above a given threshold \cite{ex1}. Each spike is associated to a precise orbit in the phase space that do not depend on the detailed nature of the stimulus and during which the system is insensitive to any new, above-threshold, perturbation. If the rest state is a fixed point or a (sub-threshold) limit cycle, excitable pulses can only be observed by adding external perturbations, or in presence of noise. On the other hand, a chaotic rest state might sporadically overcome the threshold and trigger the excitable response. The result will be an erratic--though fully deterministic--sequence of spikes on top of a chaotic background.

Although similar sequences have been recently reproduced in a semiconductor laser with opto-electronic feedback \cite{tito}, is still lacking an experimental evidence of such dynamics in the originally considered system,  that involves peculiar physical characteristics: in particular, the multiple timescale competition between dissipative (photothermal) and reactive (radiation pressure) nonlinearities. In this work we report an experimental study of such phenomena.

\begin{figure}
\begin{center}
\includegraphics*[width=1.0\columnwidth]{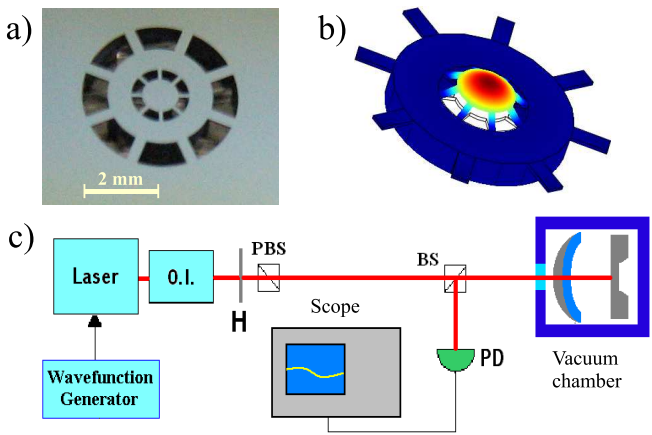}
\end{center}
\caption{a) Image of the double wheel oscillator. b) FEM calculation of the fundamental mode. c) Basic scheme of the experimental apparatus. O.I.: optical isolator; H: half-wave plate; PD: photodiode; PBS: polarizing beam-splitter; BS: beam-splitter.}
\vspace{-.3cm}
\label{figu1}
\end{figure}

Our oscillating mirror, realized on a 500~$\mu$m silicon substrate, consists of a double wheel (see Fig. \ref{figu1}a,b). The central mass (main oscillator) and the arms are 70~$\mu$m thick, while the intermediate ring mass (isolating stage) is 500~$\mu$m thick. On the front side of the wafer, a deposition of alternate Ta$_2$O$_5$/SiO$_2$ quarter-wave layers provides the highly reflective coating.
The main oscillator is used as end mirror of a $L=12.2$~mm long Fabry-Perot cavity with a $50$~mm radius silica input mirror (transmissivity $\mathcal{T}$=110 ppm) operating in a vacuum chamber at $5\cdot10^{-4}$~Pa. We obtain a cavity finesse of $\mathcal{F}$ = $3\cdot10^{4}$ (half linewidth $\gamma = 225$~kHz), with an intracavity optical power $P_c=1.1\cdot10^{4}P_{in}$, where $P_{in}$ is the optical power at the input mirror.
The mechanical characteristics of the micro-mirror has been measured in a previous experiment \cite{noi2}. We find a mechanical frequency $\omega_M / 2 \pi \sim$~250~kHz, a mechanical quality factor $Q \sim$~5~$10^{3}$ and an effective mass $m \sim ~10^{-7}$~Kg.

The aim of the experiment is to explore the system dynamics as the detuning between laser and cavity resonance, $\Delta \nu$, and the injected optical power are varied. To this end, the laser is free-running in order to avoid possible modifications of the dynamics induced by the frequency locking servo-loop.
The experimental setup is sketched in Fig.~\ref{figu1}c. The light source is a cw tunable Nd:YAG laser operating at $\lambda$=1064~nm. After a 40~dB optical isolator, the beam is injected into the cavity. The radiation intensity is controlled by an half-wave plate followed by a polarizing beam splitter and the reflected signal beam is monitored by the photodiode PD.

\begin{figure}
\begin{center}
\includegraphics*[width=1.0\columnwidth]{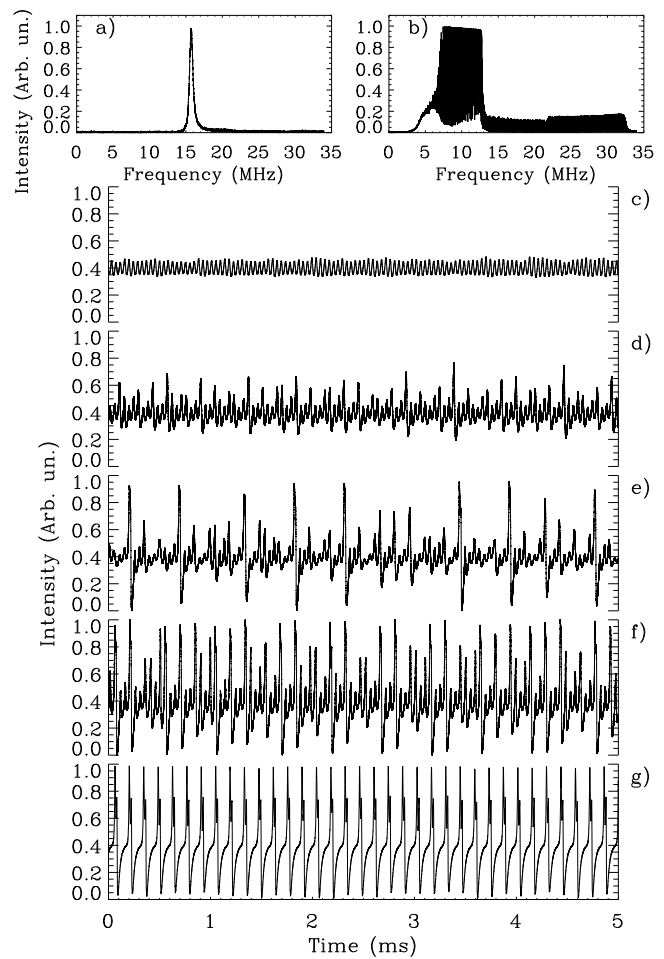}
\end{center}
\caption{Normalized reflected signal as the laser frequency is varied (scanning rate 6.8 MHz/ms) corresponding to a) $P_{in}$=0.5 mW and b) $P_{in}$=50 mW. Temporal evolution of the reflected intensity for $P_{in}$=50 mW: c) $\Delta \nu =-3.9$MHz, d) $\Delta \nu =-3.64$MHz, e) $\Delta \nu =-3.16$MHz, f) $\Delta \nu =-3$MHz, g) $\Delta \nu =-2.6$MHz. Detuning values have been estimated from the optical spectrum in Fig. \ref{figu2}b.}
\vspace{-.3cm}
\label{figu2}
\end{figure}

In Fig. \ref{figu2} we report the reflected intensity as the laser frequency is scanned starting from the red side of the cavity resonance. For simplicity, the intensity signals are normalized in order to be zero out of resonance and equal to 1 at resonance.
For low input power we observe the typical reflection dip (Fig. \ref{figu2}a). For high powers, instead, the line is strongly asymmetric and dynamical instabilities arise at both sides of the cavity resonance (Fig. \ref{figu2}b). In the high-frequency part (blue side), the system displays radiation-pressure-induced oscillations with the period of the mechanical oscillator, evolving from sinusoidal to chaotic as input power increases. This phenomenology has been previously observed in toroidal microcavities \cite{in6}.
On the other hand, on the red side of the resonance the system exhibits a complex sequence of regimes, which have never been observed in such devices. In particular, we stress that the dynamics is much slower than the natural mechanical oscillation and substantially different from that observed in Ref. \cite{in7} where it is reminiscent of a chaotic damped driven pendulum.
Figure \ref{figu2}c-g shows five traces of the reflected intensity as the
laser frequency is delicately increased, approaching the resonance from the red side. The stationary intensity value loses stability through a supercritical Hopf bifurcation and a small-amplitude quasi-harmonic limit cycle is observed (\ref{figu2}c). By further decreasing $\vert \Delta \nu \vert$, a series of bifurcations occurs and the quasi-harmonic limit cycle eventually converts into a small amplitude chaotic attractor (\ref{figu2}d). Although the noise present in the system prevents the observation of the full bifurcations sequence, the model that we will later introduce suggests that the first Hopf bifurcation is followed by a period-doubling route to chaos. As the cavity resonance is further approached a chaotic spiking regime is observed, where the small amplitude chaotic oscillations are sporadically interrupted by large pulses (\ref{figu2}e). The mean spikes rate increases as $\vert \Delta \nu \vert$ decreases (\ref{figu2}f) until the so-called relaxation oscillation regime \cite{ex1} is finally reached (\ref{figu2}g).
Note that the quasiharmonic cycle and the chaotic oscillations are clearly faster than relaxation oscillations. Moreover, the large pulses in the chaotic spiking regime have the same amplitude and roughly resemble in shape those of the relaxation oscillations regime. Indeed, accordingly to \cite{noi}, each spike should, at least partially, correspond to a single relaxation oscillation cycle (i.e., by definition an excitable pulse), triggered by sufficiently strong chaotic fluctuations.

The scenario described above is found in a detailed physical model of a
suspended mirror resonator in the presence of radiation pressure and photothermal effects. While radiation pressure increases the cavity length by pushing the mirrors, the photothermal effect decreases it due to their thermal expansion. Therefore, by injecting an input field detuned to the red from the cavity resonance, the first effect tends to increase the intracavity optical power in a characteristic time given by $2 \pi /\omega_M$ and the second one to reduce it in a time $ 2 \pi /\omega_c$, where $\omega_c$ is the photothermal response angular frequency  \cite{Braginsky}.
For low mechanical quality factors, such competition produces excitability and relaxation oscillations in the optical intensity, through a simple Hopf bifurcation \cite{noi1}. On the other hand, when the damping rate $\omega_M / Q$ becomes comparable to $\omega_c$, the transitional regime exhibits complex bifurcation sequences whose interplay with the excitable dynamics leads to a chaotic spiking regime as predicted in Ref. \cite{noi}. 

At difference with the model analyzed there, in this case the optical field evolves on a timescale comparable to that of radiation pressure and cannot be adiabatically eliminated. The model thus requires to include also the intracavity field equation and reads
\begin{eqnarray}
\dot{\varepsilon} = -[1 - i (\delta_0 + \phi + \theta)] \varepsilon + 1 \; , \label{eq1} \\
\ddot{\phi} + \frac{\omega}{Q} \dot{\phi} + \omega^2 \phi = \alpha \vert \varepsilon \vert^2 \; , \label{eq2} \\
\dot{\theta} = -\omega_{PH} (\theta + \beta \vert \varepsilon \vert^2)  \; , \label{eq3}
\end{eqnarray}
where $\varepsilon$ is the intracavity field normalized to its resonant value; $\phi$ and $\theta$ are the cavity length changes due to radiation-pressure and photothermal expansion, respectively, normalized to $\gamma_l=\gamma \frac{\lambda L}{c}$; $\delta_0$ is the detuning between laser and cavity resonance for zero input power, normalized to $\gamma$;
$\omega = \omega_M / \Gamma$ and $\omega_{PH} = \omega_c / \Gamma$ are the dimensionless oscillator and photothermal frequencies, normalized to the cavity decay rate $\Gamma$. The parameters $\alpha = \frac{2 P_c}{m c \Gamma^2 \gamma_l}$ and
$\beta$ measure the strength of the radiation pressure and photothermal \cite{marino2007} effects. Using the opto-mechanical parameters given before, we calculate $\omega=1.2$ and $\alpha=4.9~10^{-2}P_{in}$, with $P_{in}$ expressed in mW.
\begin{figure}
\begin{center}
\includegraphics*[width=1.0\columnwidth]{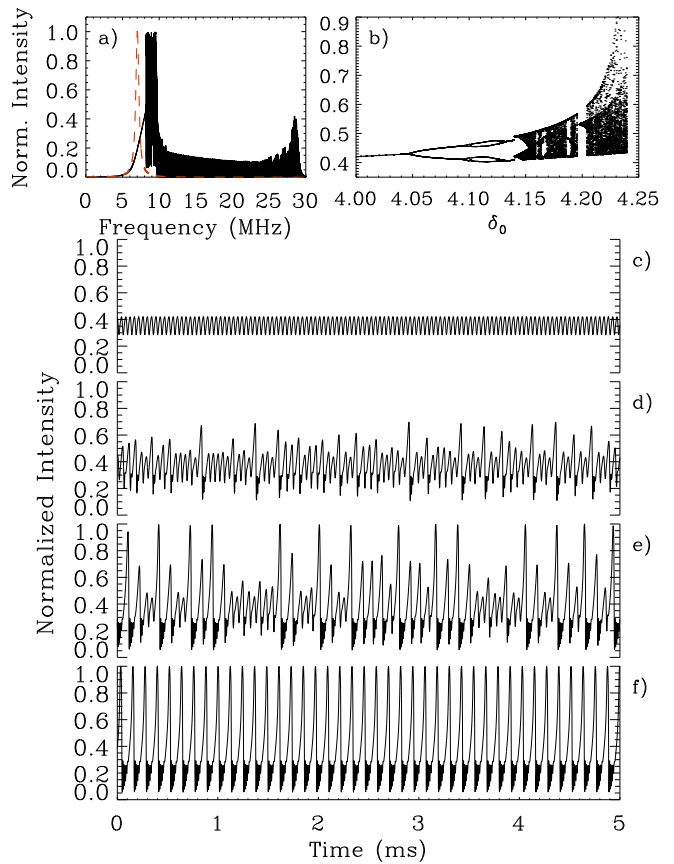}
\end{center}
\caption{Numerical integration of Eqs. (\ref{eq1})-(\ref{eq3}), with $\omega=1.2$, Q=5000, $\alpha=4.9~10^{-2}P_{in}$, $\omega_{PH}=2.4~10^{-4}$ and $\beta=3.5~10^{-1}P_{in}$, with $P_{in}$ expressed in mW.
a) Normalized reflected intensity $I = \vert \varepsilon \vert^2$ as $\delta_0$ is linearly varied in time, with a rate of 6.8 MHz/ms, during the numerical integration; $P_{in}$=0.5 mW (red dashed line) and $P_{in}$=50 mW (black solid line). b) Bifurcation diagram for the peak values of $I$ ($P_{in}$=50 mW). c) Time traces of $\vert \varepsilon \vert^2$ for $P_{in}$=50 mW: c) $\delta_0=4$, d) $\delta_0=4.22$, e) $\delta_0=4.25$, f) $\delta_0=4.29$.}
\vspace{-.3cm}
\label{figu3}
\end{figure}
Concerning the photothermal effect, here we remark that Eq. (\ref{eq3}) realistically describes the photothermal dynamics only at frequencies well above $\omega_{PH}$. Indeed, the frequency response of the length variations to a modulated intracavity power has a slow, logarithmic divergence at low frequencies ~\cite{cerdonio,marin}, while it follows an intuitive 1/frequency-dependence as in a standard single-pole system at high frequencies. The photothermal response is dominated by the fused silica input mirror for which we calculate $\omega_c/ 2 \pi \sim$~50~Hz, leading to $\omega_{PH}=2.4~10^{-4}$, and $\beta=3.5~10^{-1}P_{in}$ given an absorption of 2 ppm \cite{marino2007,Braginsky}.

Fig. \ref{figu3}a shows the reflected intensity signal as the laser frequency is varied. The signal has been obtained by numerical integration of Eqs. (\ref{eq1})-(\ref{eq3}) with $\delta_0$ linearly varying in time. As in the experiment, for high powers the line shape becomes strongly asymmetric with dynamical instabilities at both sides of the resonance. We remark, however, that the line broadening and the width of the instability region on the red side are quite different from the experiment. This is due to the single-pole description of the photothermal effect that, for such slow scanning rates, is no longer a good approximation.
On the other hand, the essential features of the chaotic spiking regime, which is significantly faster, are well reproduced. In particular, the initial Hopf bifurcation takes place when the mean normalized intracavity intensity is approximately equal to $0.4$ (see Fig. \ref{figu3}c). Then, the system undergoes a period doubling cascade leading to the birth of small-amplitude chaotic attractors (Fig. \ref{figu3}d). This is illustrated by the bifurcation diagram in Fig. \ref{figu3}b.
Further increasing $\delta_0$, the mean amplitude of the attractors grows, until that chaotic fluctuations are sufficiently large to eventually trigger excitable spikes (approximately at $\delta_0$=4.23). The result is an erratic -- sensitive to initial conditions -- sequence of pulses on top of a chaotic background (Fig. \ref{figu3}e). For larger $\delta_0$, the mean firing rate increases until a periodic regime is finally reached.
\begin{figure}
\begin{center}
\includegraphics*[width=1.0\columnwidth]{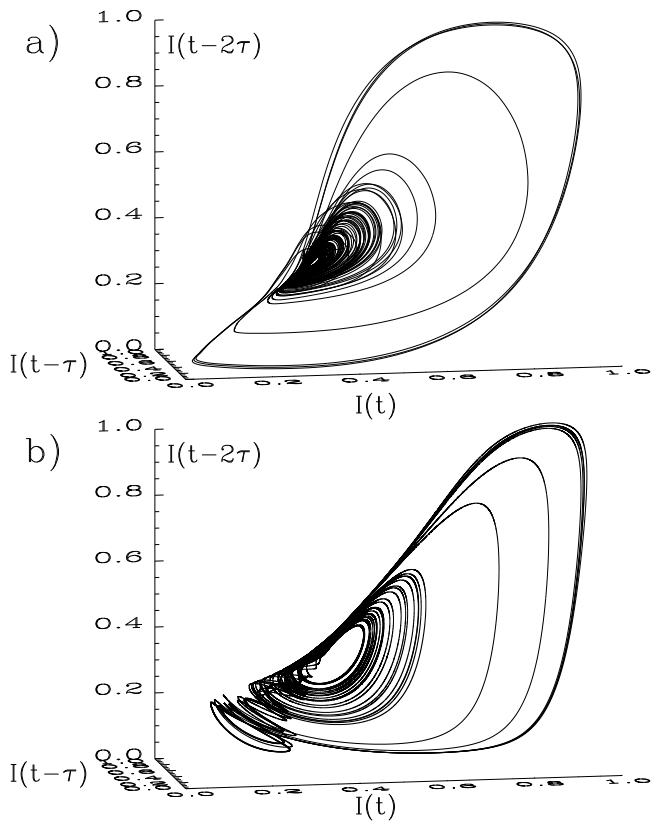}
\end{center}
\caption{Three-dimensional reconstruction of the phase-portrait through the Ruelle-Takens embedding technique, from a) experimental and b) numerical time-series of the optical intensity in the chaotic spiking regime (parameters as in Fig. \ref{figu2}e and Fig. \ref{figu3}e). In both cases we have used a time delay $\tau$=1.74 $\mu$s.}
\vspace{-.3cm}
\label{figu4}
\end{figure}
In Fig. \ref{figu4}a we plot a three-dimensional phase-space projection of the chaotically spiking attractor reconstructed from experimental time series by means of the Ruelle-Takens embedding technique.
This reconstruction shows all the essential features of the phase portrait reported in \cite{noi} where the small-amplitude chaotic attractor sporadically triggers large excursions in the phase space. These orbits, associated to a spike in the time series, are always approximately equal one to each other.
Fig. \ref{figu4}b shows the same reconstruction obtained from numerical data showing a good qualitative agreement.

In conclusion, we have shown that the multiple-time scale dynamics induced by radiation pressure and photothermal effects in suspended mirror resonators, leads to
novel dynamical scenarios where a small-amplitude chaotic attractor spontaneously generates irregular spike sequences.
The dynamical mechanism at the basis of this phenomenon is relevant also in biology, for instance in the study of some apparently erratic neural bursting that might have well defined coding functions.
Our resonator, operating at cryogenic temperatures and with slightly adjusted oscillator parameters, is expected to achieve a regime in which quantum fluctuations of radiation pressure are the dominant noise source \cite{noi2}, and eventually the oscillator is close to its quantum ground state. The instability here investigated, while making the observation of quantum-mechanical properties of the system more tricky, might enable the study of the transition from classical to quantum domain of a chaotic system \cite{folclore}.

We thank F.S. Cataliotti and M. Inguscio for valuable discussion and careful reading of the manuscript.

\end{document}